\documentstyle[sprocl,amstex,epsfig]{article}

\def\be{\begin{equation}}
\def\ee{\end{equation}}
\def\bea{\begin{eqnarray}}
\def\eea{\end{eqnarray}}
\begin{document}

\title{``Good-Walker'' + QCD dipoles   = Hard Diffraction\footnote{Invited talk 
 at the DIS98 workshop, Brussels, April 1998.}
}

\author{R. Peschanski}

\address{Service de Physique Th\'eorique, CEA, CE-Saclay\\
F-91191 Gif-sur-Yvette Cedex, France}


\maketitle\abstracts{The Good-Walker mechanism for diffraction
is shown to provide a link between total and diffractive structure functions and
to be relevant for QCD calculations at small $x_{Bj}.$ For
Deep-Inelastic scattering on a small-size target (cf. an onium) the 
r\^ ole of Good-Walker ``diffractive eigenstates'' is played by the QCD dipoles appearing in the $1/N_C$ limit of QCD. Hard diffraction is thus related to the QCD tripe-dipole vertex which has been recently identified (and
calculated) as being a conformal invariant correlator and/or a closed-string amplitude. An extension to hard diffraction at HERA via $k_T-$factorisation of the proton vertices leads to interesting phenomenology.}

\section{The Good-Walker mechanism:}

The Good-Walker mechanism \cite{good} is known to provide a simple explanation of the link between two phenomena of high-energy (soft) scattering: {\it absorption} and {\it diffractive dissociation}. Our aim is to show that the
mechanism can be used in QCD calculations of hard scattering at small $x_{Bj}$ providing a simple
connection between total and diffractive structure functions.

Let us consider the diffraction of a set of quantum states on a potential. {\it Absorption} describes the absorption of a given initial state due to the presence of inelastic channels. {\it Diffractive dissociation} is the observation that there exists transition between different
such states, i.e. the transition matrix between initial and final diffractive
states is not diagonal {\it a priori}. The Good-Walker mechanism \cite{good}
shows that the two phenomena are related through the fluctuations of the absorption factors. Let $<\!i|t|i\!>$ be the diffractive amplitude of a given
initial state and consider a orthonormal diagonal basis $|m\!>$ of the transition matrix
we can write :
\begin{equation}
<\!i|t|i\!> \equiv \sum_{m,n}<\!i|m\!><\!m|t|n\!><\!n|i\!> = \sum_{m}\mid<\!i|m\!>\mid^2<\!m|t|m\!>\Rightarrow \bar t,
\label{2}
\end{equation}
where $\bar t$ is the average absorption factor. With the same notations, we may write diffraction-dissociation cross-sections in term of:
\begin{equation}
\sum_{m} <\!i|m\!><\!m|\ t t^{\dag}\ |m\!> <\!m|i\!>  - <\!i|t|i\!> <\!i|t^{\dag}|i\!>\Rightarrow  \overline {t\ t^{\dag}  } -  \bar t\ \bar t^* .
\label{4}
\end{equation}
From formulae (1,2) it becomes clear that the total contribution
of diffractive dissociated states is related to the dispersion over  absorptive factors. In the case of ``soft'' diffraction, these formulae relate total
and diffractive cross-sections (actually for each impact parameter or partial
wave). As we shall now see in ``hard diffraction'', it is a convenient way to relate total and diffractive structure functions and apply QCD calculations
 at small $x_{Bj}$ to both observables. .

\section{Hard diffraction off a {\it hard} target}

In the past, there were interesting attempts \cite{miettinen} to identify the diagonal basis,
or {\it  diffractive eigenstates} with free partons. However, the applications
to ``soft'' reactions prevent from the use of perturbative QCD calculations.
On the other hand, partons (gluons) are not necessarily diffractive eigenstates
 in high-energy (small $x_{Bj}$) processes. In a recent approach\cite{bia1,bia2}, it was suggested to use the QCD dipole states as the diagonal basis of diffractive eigenstates in a hard scattering process, see
fig.1. QCD dipole states appear \cite{muel1} in the  $1/N_C$ limit of QCD at high energy. The key observation is that the QCD dipole states interact purely elastically by the exchange of two gluons. On the other hand, the wave function of initial hard
$q \bar q$ (onium) states in terms of interacting dipoles is known from QCD
calculations in the infinite momentum frame \cite{muel1}. Thus both   the matrix elements $<\!m|t|m\!>$ and the coefficients $<\!i|m\!>$ in formulae (1,2) are determined in a suitable  perturbative
QCD framework.

In order to apply these properties to structure functions, let us consider
the (theoretical) process of Deep Inelastic Scattering (DIS) on an onium target. In the same spirit as the Good-Walker derivation, two different
components to hard diffraction happen to be relevant, the {\it elastic} and
{\it inelastic} components corresponding to, respectively, the elastic and 
dissociative diffraction previously considered in soft processes.  The virtual photon is represented by a well-defined \cite {nikzak} set of $q \bar q$  initial states which give rise to a collection of QCD interacting dipoles following \cite{muel1}. The interaction of  QCD dipoles
from the photon with those from the onium give rise to a total structure function given by BFKL dynamics \cite{BFKL}. In the QCD dipole picture of the Good-Walker mechanism, it amounts to compute the distribution of absorptive factors as a function of the QCD dipoles, in practice  as a function of their transverse sizes. Considering  the inelastic component, one   investigates 
 the simultaneous interaction of {\it two} dipoles from the photon, see Fig.1a,  to compute the dispersion of dipole sizes, and thus to generate a significant contribution to diffractive dissociation \cite{bia1}.

\begin{figure}[t]
\begin{center}
\mbox{\epsfig{figure=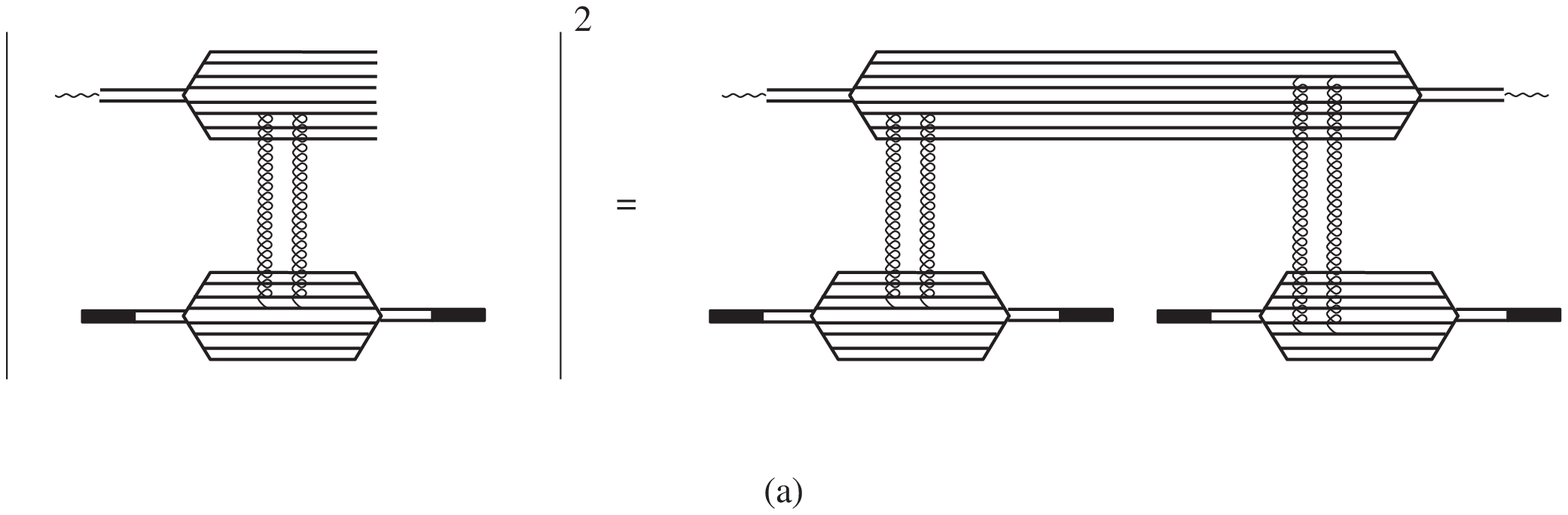,height=8.5cm}}

\mbox{\epsfig{figure=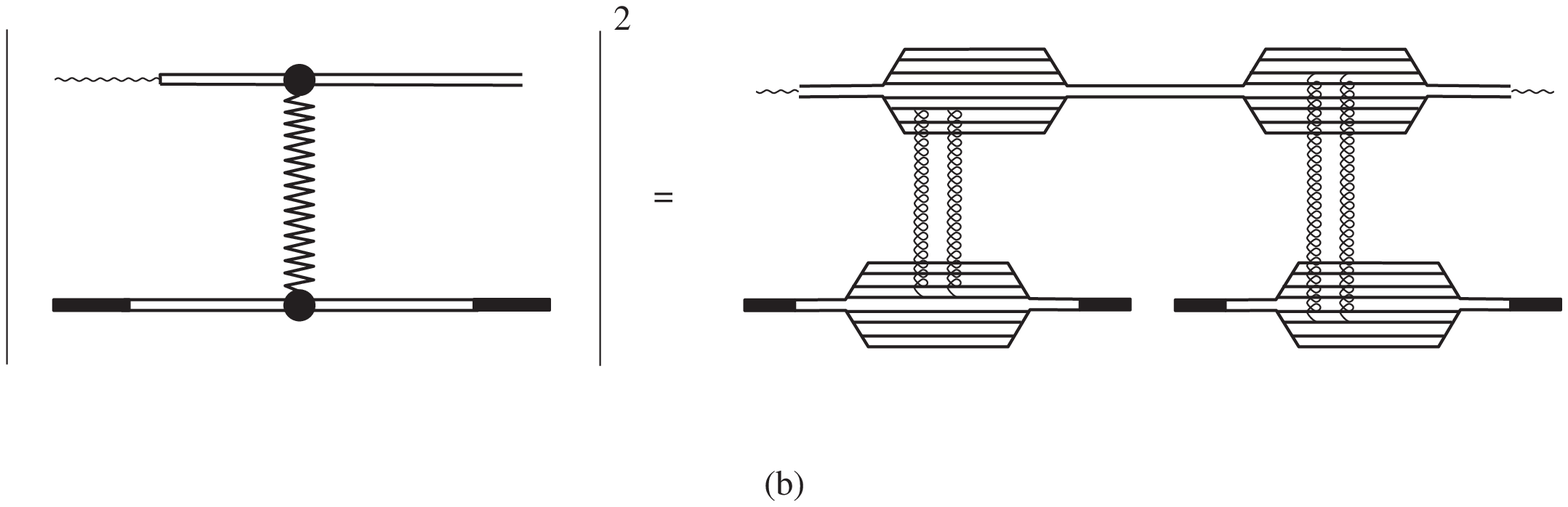,height=8.5cm}}
\vspace{-0.5cm}
\label{fig:fig1}
\caption{{\bf the two diffractive components}
(a) Inelastic (b) Elastic}.
\end{center}
\end{figure}

There is however a distinct component \cite{nikzak} which is analoguous to elastic diffraction in soft reactions. In this process, see Fig.1b, the photon $q \bar q$ states interact elastically with the target. The calculation of this component with QCD dipoles has been performed \cite{bia2} and requires a novel
quantum-mechanical aspect of QCD dipole calculations lying beyond the original Good-Walker description. Indeed, while the derivation has been made for the 
total cross-sections (eventually for each impact-parameter), it cannot be used 
for  a given mass $M$ of the diffracted state
(neither for a given rapidity gap $\approx Y - \log M^2$ between the diffractive state and the target). In fact one cannot  diagonalize
the momentum operator and thus the mass of the diffractive state on the QCD
dipole basis since \cite{muel1}  the QCD dipole basis requires
kinematics to be described in a mixed representation using transverse coordinates and rapidity and not full momentum space. The correct implementation of this effect leads to interference
terms in the final formulae \cite{bia2}.

Interestingly enough,  diffractive processes happen to be related to some quite fundamental theoretical aspects of  (resummed) perturbative QCD. The inelastic component has been shown \cite{bia3} related to the $1 \to 2$ dipole transition vertex which, in turn, has a string theory interpretation \cite{pesch} in terms of a closed string amplitude with 6 legs.
It is possible to explicitely compute the triple-dipole vertex which 
appears to be quite large \cite{bia4}. The computation of the same quantity can be done also in the framework of conformal field theory \cite{kor}. The stringy character of     any $1 \to n$ dipole transition vertex \cite{pesch} may lead to interesting theoretical developments. The other
(elastic) component can also be exactly computed \cite{bia5} with the help of
a  derivation \cite {navelet} of the conformal coupling of a BFKL Pomeron to a general $q \bar q$ state.

\section{Hard diffraction off a {\it soft} target}

The application of the QCD dipole formalism to a more realistic target,
e.g. a proton of the HERA ring, requires some care and simplifying assumptions. Indeed there exists large theoretical uncertainties in the use
of perturbative QCD for hard scattering on a ``soft'' target. Let us briefly mention some of them. $k_T$-diffusion of the intermediate gluons \cite{bartels}
lead to a excursion inside the strong coupling domain of QCD near the proton vertex. More recently, a rather stringent upper limit in $x_{Bj}$    due to the breaking of the Operator Product Expansion has been reported \cite {muel2}. At the present conference were reported for the first time  calculations of large next-to-leading BFKL corrections \cite {BFKL2} which may invalidate predictions for proton structure functions. It is tempting to relate these theoretical objections to an old idea of Bjorken \cite{Bjorken}. From the calculation of the photon wave function \cite{nikzak} it appears that the effective virtuality of the 
photon $q \bar q$ state is not $Q^2$ but ${\hat Q}^2 = z (1-z) Q^2,$ where z is the momentum fraction of the photon beared by the quark. Thus, if the probability of $z$ (or $1-z$) being small is sizeable, the effective virtuality
may be  of the same order of that of the target. In that case, the process may indeed be soft and thus not governed by perturbative calculations.

However, some arguments may be opposed to such a skepticism from both
theoretical and phenomenological sides. It has been argued at this conference
\cite{muel3} that the same quantum state configurations which may invalidate 
a perturbative treatment of diffractive scattering may be washed out by 
the inclusive QCD resummation of structure functions. Moreover the ``soft'' part of the cross-section may be eaten out by the strong absorption expected from soft diffractive 
components. On a more phenomenological ground, there are hints that QCD dipole
descriptions of proton structure functions do agree with present data
using a small number of parameters describing the 
non-perturbative proton input \cite {navelet2}. Indeed, the $k_T$-factorization property of high-energy QCD \cite{catani} may be invoked to relate different structure functions of the proton  \cite {munier}. Extending this factorization property to diffraction dissociation at the photon vertex, it is possible to find a convenient and economical (in terms of parameters)
description of diffractive structure functions \cite {royon}.

\section*{Acknowledgments}

The material of this contribution is the result of  a long term and stimulating collaboration with Andrzej Bialas, Henri Navelet and Christophe Royon.
This work was supported in part by the EU Fourth Framework Programme `Training 
and Mobility of Researchers', Network `Quantum Chromodynamics and the Deep Structure 
of Elementary Particles', contract FMRX-CT98-0194 (DG 12 - MIHT).

\end{document}